\documentclass[reprint, 
amsmath,amssymb,
aps,prd,
]{revtex4-2}

\usepackage{graphicx}
\usepackage{float}  %
\usepackage{bm}
\usepackage{booktabs}
\usepackage{amsmath}
\usepackage{mathrsfs}
\usepackage{subfigure}
\usepackage{enumerate}

\setcitestyle{numbers,maxcitenames=3}

\usepackage[hidelinks]{hyperref}
\hypersetup{
    colorlinks=true,
    linkcolor=blue,
    filecolor=gray,      
    urlcolor=blue,
    citecolor=blue,
}

\allowdisplaybreaks[1]
            
\begin{document}


\title{Inspiral waveforms of charged compact binaries and observational constraints}

\author{Zi-Han Zhang$^{1,2}$}
 \email{zhangzihan242@mails.ucas.ac.cn}

\author{Wen-Hong Ruan$^{3,4}$}
  \email{ruanwenhong@hunnu.edu.cn}
 
\author{Tan Liu$^{5}$}
 \email{lewton@mail.ustc.edu.cn}

\author{Zong-Kuan Guo$^{6,7,8}$}
\email{guozk@itp.ac.cn}

\affiliation{$^{1}$International Centre for Theoretical Physics Asia-Pacific, University of Chinese Academy of Sciences, 100190 Beijing, China}
\affiliation{$^{2}$Taiji Laboratory for GW Universe, University of Chinese Academy of Sciences, 100049 Beijing, China}

\affiliation{$^{3}$ Department of Physics, Key Laboratory of Low Dimensional Quantum Structures and Quantum Control of Ministry of Education, and Institute of Interdisciplinary Studies, Hunan Normal University, Changsha, 410081, China}

\affiliation{$^{4}$ Hunan Research Center of the Basic Discipline for Quantum Effects and Quantum Technologies, Hunan Normal University, Changsha, 410081, China}

\affiliation{$^{5}$School of Physics and Optoelectronic Engineering, Yangtze University, Jingzhou 434023, China}

\affiliation{$^{6}$School of Fundamental Physics and Mathematical Sciences, Hangzhou Institute for Advanced Study, University of Chinese Academy of Sciences, Hangzhou 310024, China}

\affiliation{$^{7}$School of Physical Sciences, University of Chinese Academy of Sciences, Beijing 100049, China }

\affiliation{$^{8}$Institute of Theoretical Physics, Chinese Academy of Sciences, Beijing 100190, China }


\date{\today}

\begin{abstract}
Electric charges carried by compact objects can affect binary dynamics and imprint characteristic signatures on gravitational-wave signals. We derive next-to-leading-order post-Newtonian frequency-domain waveforms for charged compact binaries in both the gravitational-quadrupole and electric-dipole dominated regimes, including electromagnetic corrections beyond leading-order electric-dipole radiation that reduce the degeneracy between the component charge-to-mass ratios. Using gravitational wave events from GWTC-5.0, we place constraints on the charge-to-mass ratio and mass of each component of the binary system.

\end{abstract}

\maketitle
 
\section{Introduction}
The direct detection of gravitational waves (GWs) from compact-binary coalescences has opened a new avenue for testing  the nonlinear dynamics of gravity in the strong-field regime~\cite{PhysRevLett.116.061102,PhysRevLett.116.241103,PhysRevLett.118.221101,PhysRevLett.119.141101,PhysRevLett.125.101102,PhysRevX.13.041039}. 
For binary black holes (BBHs), the observed signals are well described by general relativity, in which the exterior spacetime of an isolated, stationary, asymptotically flat black hole is specified by its mass, spin, and charge. 
Astrophysical black holes are usually expected to be nearly neutral because ordinary electromagnetic (EM) charges are efficiently discharged by the surrounding plasma and quantum pair-production processes~\cite{10.1093/mnras/stz1904,PhysRevLett.35.463,Gibbons1975,Cardoso_2016,sym15020537}. 
Despite existing constraints on binary charge-to-mass ratios~\cite{Wang2021,PhysRevD.109.024058,PhysRevLett.126.041103}, the possibility of small residual charges remains phenomenologically interesting.

The effects of charge on compact-binary dynamics have been widely studied within Einstein–Maxwell theory.  {Numerical-relativity simulations of the inspiral and merger of charged black holes, together with merger-ringdown analyses, have further shown that existing GW observations can place meaningful bounds on the charge-to-mass ratio \cite{PhysRevLett.126.041103,PhysRevD.109.024058}.} It has also been extensively studied in the post-Newtonian (PN) framework ~\cite{Maggiore:2007ulw,Gravity_PoissonWill,PhysRevD.108.024020,placidi2025chargedblackholebinaryevolution, PhysRevD.98.104010,tang2025inspiralingbinarychargedblack,verma2026postnewtoniandynamicsradiatingcharges,alonzoartiles2026reissnernordstromblackholessecond}. In the inspiral regime, the Coulomb interaction affects the conservative dynamics, while the accelerated charges generate EM radiation in addition to the gravitational radiation.  { The conservative dynamics has been derived up to 2PN order, and the leading -1PN dissipative dynamics has been obtained, with the electromagnetic effects characterized by the charge-to-mass ratios $\kappa_1$ and $\kappa_2$ of the compact components~\cite{PhysRevD.109.084048,PhysRevD.102.103520,Liu2021}.  In our previous work \cite{zhang2026postnewtoniandynamicschargedcompact},  the completely 0PN terms and a part of 1PN terms of dissipative part have been derived. We have obtained the orbital-frequency evolution for quasi-circular inspirals, which includes the next-to-leading EM effects.}
The leading EM dipole flux enters the Fourier phase at the formal -1PN order, and can therefore be especially important at low frequencies \cite{Cardoso_2016,Christiansen_2021,Wang2021}.  {The combination $|\kappa_1-\kappa_2|$ of charge-to-mass ratios has been constrained by Bayesian inference using LIGO–Virgo observations \cite{Wang2021}.} These studies indicate that waveform models including charge effects are useful not only for testing black-hole electrovacuum physics, but also as probes of hidden-sector interactions.

In this paper, we  {continue the effort to study the motion and radiation of charged binaries.} We  extend the previous PN analysis to construct GW waveforms for charged compact binaries in the quasi-circular orbits. We first compute the time-domain polarizations by matching the radiative multipole moments to the source multipoles with the EM corrections. Following the TaylorF2 construction, we derive the frequency-domain waveform using the stationary-phase approximation and include EM corrections  in the Fourier phase. Finally, we combine our electromagnetic corrections with the standard 3.5PN spinning phase and perform Bayesian parameter estimation for three selected O4b BBH events in GWTC-5.0 \cite{theligoscientificcollaboration2026gwtc50observationssecondfourth}: GW240925\_005809, GW241102\_124058, and GW250119\_190238. This analysis places separate constraints on $\kappa_1$ and $\kappa_2$ and examines their degeneracy.The Bayesian analysis yields posterior medians and 68\% credible intervals for the charge-to-mass ratios of the components $\kappa_1$ and $\kappa_2$, and $\mathcal{Q}=\kappa_1\kappa_2$, the component masses $m_1$ and $m_2$, and the chirp mass $\mathcal{M}_c$. These results connect the analytic charged-binary waveform model with observational constraints on possible electric charges.

This paper is organized as follows. 
In Sec.~\ref{Sec2} we derive the time-domain polarizations of charged compact binaries from the multipolar post-Minkowskian waveform and illustrate the effect of charge on the inspiral time-domain waveform. 
In Sec.~\ref{Sec3} we construct the frequency-domain waveform in both the gravitational-quadrupole and electric-dipole dominated regimes and analyze the PN hierarchy of the EM phase corrections. In Sec. \ref{Sec4} we perform Bayesian inference using three selected GW events to constrain the component charge-to-mass ratios and masses. Throughout this work, $G$ denotes the gravitational constant and $c$ denotes the speed of light.

\section{Time-Domain waveforms}\label{Sec2}
\subsection{Waveform corrections of charged binaries}
In radiative coordinates $(T, \bm{X})$, the general multipolar-post-Minkowskian solution for the transverse-traceless (TT) GW field can be written as \cite{Blanchet2024,Gravity_PoissonWill}
\begin{align}
H_{ij}^{\mathrm{TT}}(U,\bm X)=&\frac{4G}{c^2 R}\,\mathcal P_{ijab}(\bm N)\sum_{\ell=2}^{+\infty}\frac{1}{c^\ell \ell!}\bigg[N_{L-2} \mathrm{U}_{abL-2}\notag\\
-&\frac{2\ell}{c(\ell+1)}
N_{cL-2}
\epsilon_{cd(a} \mathrm{V}_{b)dL-2}
\bigg]+\mathcal{O}\left(R^{-2}\right).
\end{align}
where $R$ is the distance from the GW source to the detector and $\epsilon$ is the Levi-Civita symbol.  {This result is also available for the charged binaries.} $\mathcal{P}_{ijab}=\mathcal{P}_{ia}\mathcal{P}_{jb}- \mathcal{P}_{ij}\mathcal{P}_{ab}/2$ is the TT
projection operator, with $\mathcal{P}_{ij}=\delta_{ij}-N_iN_j$ being the projector onto the plane orthogonal to $N_i=(\bm{N})_i=(\bm{X})_i/|\bm{X}|$. Here $U \equiv T-R/c$ is the retarded time,
$\bm N \equiv \bm X/R$, and $
N_{L-2}=N_{i_1}\cdots N_{i_{\ell-2}}$. At leading post-Minkowskian order, we match the radiative multipole moments to the corresponding source multipole moments,
\begin{equation}
\mathrm{U}_L=\mathrm{I}^{(\ell)}_L+\mathcal{O}(G),\quad \mathrm{V}_L=\mathrm{J}^{(\ell)}_L+\mathcal{O}(G),
\end{equation}
We obtain
\begin{align}\label{HTT}
    H^{\rm TT}_{ij}(U,\mathbf X)
=&
\frac{2G}{c^4R}\,
\mathcal P_{ijab}(\bm{N})
\bigg[
I_{ab}^{(2)}\notag\\
+&\frac{1}{c}
\left(
\frac{1}{3}N_c I_{abc}^{(3)}
-\frac{4}{3}N_c\epsilon_{cd(a}J_{b)d}^{(2)}
\right)\notag\\
+&\frac{1}{c^2}
\left(
\frac{1}{12}N_cN_d I_{abcd}^{(4)}
-\frac{1}{2}N_cN_d\epsilon_{ce(a}J_{b)ed}^{(3)}
\right)
\bigg]\nonumber\\[6pt]
+&\mathcal O \left(R^{-1}c^{-7},R^{-2}\right) .
\end{align}

 Using the transformation between radiation coordinate $(T,X^i)$ and harmonic coordinate $(t,x^i)$, we obtain $N^i=n^i+\mathcal{O}(r^{-2})$ with $n^i=x^i/r$ and $r=|x^i|$. In Eq. \eqref{HTT}, all contributions are evaluated consistently in order $c^{-6}$. Since the dominant term is the mass quadrupole contribution, the corresponding source moment $I_{ab}$ must be known up to 1PN accuracy. The mass octupole $I_{abc}$,  hexadecapole $I_{abcd}$, the current quadrupole $J_{ab}$, and octupole $J_{abc}$, are required only at the Newtonian level. In the center-of-mass frame, the required multipole moments are \cite{zhang2026postnewtoniandynamicschargedcompact,PhysRevD.43.3259,1989MNRAS.239..845B}
\begin{align}\label{MQ}
    \hat{\mathrm{I}}^{ij}=&\mu x^{\langle i}x^{j\rangle}\notag\\
    +&\frac{\mu}{c^2}\bigg\{\bigg[\frac{29}{42}(1-3\eta)v^2-\frac{1}{7}(5-8\eta)\frac{G m}{r}\mathcal{Z}\bigg]x^{\langle i}x^{j\rangle}\notag\\
    -&\frac{4}{7}(1-3\eta)r\dot{r}x^{\langle i}v^{j\rangle}+\frac{11}{21}(1-3\eta)r^2v^{\langle i}v^{j\rangle}\bigg\}\nonumber\\[4pt]
    +&\mathcal{O}(c^{-4}),
\end{align}
where total mass $m\equiv m_1+m_2$, symmetric mass ratio $\eta\equiv m_1m_2/m^2$. For the EM part we define
\begin{equation}
    \mathcal{Z}\equiv 1-\mathcal{Q}, \qquad\mathcal{Q}\equiv\kappa_1\kappa_2,
\end{equation}
and $\kappa_i\equiv q_i/(\sqrt{4\pi \epsilon_0 G} m_i)$ for $i=1,2$. The factors of $\mathcal{Z}$ arise from the EM contribution to the energy-momentum tensor and from the Coulomb correction of the relative acceleration. The remaining mass and current multipole moments required at Newtonian order are
\begin{align}
    \hat{\mathrm{J}}^{ij}&=\mu\sqrt{1-4\eta}\epsilon^{kl\langle i}x^{j\rangle}x^k v^{l}+\mathcal{O}(c^{-2}),\\
    \hat{\mathrm{J}}^{ijk}&=\mu(1-3\eta)\epsilon^{ab\langle i}x^jx^{k\rangle}x^a v^b+\mathcal{O}(c^{-2}),\\
    \hat{\mathrm{I}}^{ijk}&=\mu\sqrt{1-4\eta}x^{\langle i}x^j x^{k\rangle}+\mathcal{O}(c^{-2}),\\
    \hat{\mathrm{I}}^{ijkl}&=\mu(1-3\eta)x^{\langle i}x^jx^kx^{l\rangle}+\mathcal{O}(c^{-2}),
\end{align}
where $\mu=m\eta$. The symbol ``$\hat{\quad}$'' denotes the symmetric trace-free (STF) tensor, $\hat{x}_{ij}=x_{\langle i}x_{j\rangle}=x_{( i}x_{j)}-\delta_{ij}x_kx^k/3$ and $x_{( i}x_{j)}=(x_ix_j+x_jx_i)/2$.

Kepler’s third law, including EM correction at 1PN order, is \cite{zhang2026postnewtoniandynamicschargedcompact}
\begin{align}
    \omega^2&=\frac{Gm}{r^3}\bigg(\mathcal{Z}+\frac{1}{c^2}\frac{Gm}{r}\mathcal K\bigg)+\mathcal{O}(c^{-4}).\label{omega}\\
    r&=\Big(\frac{Gm}{\omega^2}\mathcal{Z}\Big)^{1/3}\bigg(1+\frac13{\rm{x}}\mathcal{Z}^{-1/3}\mathcal K\bigg)+\mathcal{O}(c^{-4}),\label{ro}
\end{align}
where the leading order EM correction $\mathcal{Z}$ of Kepler's third law comes from the Coulomb interaction. We define factors
\begin{gather}
    \mathcal K\equiv\eta-3-\zeta_0+\bigg(\frac{9}{2}+\eta\bigg)\mathcal{Q}-\bigg(2\eta+\frac{1}{2}\bigg)\mathcal{Q}^2,\\
    {\zeta}_0\equiv \kappa_1^2\chi_1+\kappa_2^2\chi_2,
\end{gather}
where $\chi_1=m_1/m$ and $\chi_2=m_2/m$ are the component mass fractions. 
In the neutral limit $\kappa_1=\kappa_2=0$, factor $\mathcal{K}$ reduces to the standard 1PN result $\mathcal{K}=\eta-3$, while factor $\zeta_0$ becomes 0. The PN terms are ordered by means of the frequency-related variable
\begin{equation}\label{x}
    {\rm{x}} =(Gm\omega)^{2/3}c^{-2}.
\end{equation}
The plus and cross polarization states of
the asymptotic waveform are defined by \cite{Blanchet2024}
\begin{align}\label{h}
    h_{+,\times} = \frac{2 G \mu }{c^2 R} \mathcal  Z^{2/3}{\rm{x}}\sum_{p=0}^{2}\mathcal  Z^{p/3}{\rm{x}}^{p/2} H_{p/2}^{+,\times}(\phi, c_\iota, s_\iota),
\end{align}
where $\phi(t)$ is the orbital phase, with $\dot{\phi}=\omega t=\pi f t$, and $f$ is the GW frequency. We denote by $\iota$ ( $0<\iota<\pi$) the inclination angle between the line-of-sight unit vector $\bm{N}$
(from the source to the detector) and the orbital angular-momentum direction. We use
the shorthand $c_\iota \equiv \cos\iota$ and $s_\iota \equiv \sin\iota$ for the cosine and sine of the angle of inclination. We find that the GW strain is affected by the EM effects through the factor $\mathcal{Z}$. This dependence arises from the affected Kepler relation and the EM correction to the mass quadrupole moment in Eq. 
\eqref{MQ}.
We obtain
\begin{align}
    H_{0}^{+} =& -\left(1 + c_\iota^{2}\right) \cos 2\phi, \\
{H}^{+}_{1/2} =& s_\iota \Delta \left[ \frac{9}{8}\cos 3\phi \left( 1 + c_\iota^{2}\right)- \frac{1}{8}\cos\phi \left( 5 + c_\iota^{2}  \right) \right], \\
H^+_1=&\cos2\phi\bigg[\frac{7}{6}-\frac{5}{2}\eta-\bigg(\frac{1}{2}-\frac{5}{2}\eta\bigg)c_\iota^2\notag\\
&\qquad\quad+\frac23\mathcal  Z^{-2}\mathcal{K}(1+c_\iota^2)-\bigg(\frac{1}{3}-\eta\bigg)c_\iota^4\bigg]\notag\\
-&\cos4\phi\bigg[\bigg(\frac{4}{3}-4\eta\bigg)s_\iota^2(1+c_\iota^2)\bigg] .
\end{align}
For the cross polarizations we obtain
\begin{align}
    H_{0}^{\times} =& -2c_\iota \sin 2\phi,\\ 
H_{1/2}^{\times} =& s_\iota c_\iota \Delta \left[ -\frac{3}{4}\sin\phi + \frac{9}{4}\sin 3\phi \right], \\
H^\times_1=&c_\iota\sin2\phi\left[\frac{5}{3}-3\eta-\frac{4}{3} \mathcal Z^{-2}\mathcal K-\bigg(\frac43-4\eta\bigg)c_\iota^2\right]\notag\\
-& \sin4\phi\bigg[\bigg(\frac{8}{3}-8\eta\bigg)c_\iota s_\iota^2\bigg] .
\end{align}

Substituting the gravitational and electromagnetic radiation fluxes into the balance equation, we obtain the evolution of the orbital angular frequency  \cite{zhang2026postnewtoniandynamicschargedcompact}
\begin{align}\label{Domega}
    \dot{\omega}=&\frac{c^6\eta \mathrm{x}^{9/2}}{G^2 m^2}\bigg[2\xi_1^2+\frac{x}{\mathcal{Z}^{4/3}}\bigg(\frac{96}{5}+\sum_{a=0}^3\mathcal{C}_a\mathcal{Q}^a \bigg)\notag\\
    &+\frac{x^2}{\mathcal{Z}^{8/3}}\bigg( -\frac{264 }{5}\eta-\frac{1486}{35}+\sum_{a=0}^6\mathcal{D}_a\mathcal{Q}^a\bigg)\bigg]+\mathcal{O}(c^{-9}),
\end{align}
where the coefficients $\mathcal{C}^a$ and $\mathcal{D}^a$ are listed in Appendix \ref{App.A}. We obtain the gauge-invariant GW frequency associated with the approximate 1PN innermost stable circular orbit (ISCO), which we use as the upper frequency cutoff
\begin{align}\label{fISCO}
f_{\rm ISCO}=&\frac{c^3\mathcal{Z}^{2}}{G m \pi}\bigg[6-\mathcal{Z}\zeta_0-\frac{1}{2}\Big(15-9\mathcal{Z}+12\eta\Big)\mathcal{Q}\notag\\
&-\frac{1}{2}(1+4\eta)(-3+\mathcal{Z})\mathcal{Q}^2\bigg]^{-3/2}.
\end{align}
The ISCO radii of different charged binaries are shown in Fig. 1 of \cite{zhang2026postnewtoniandynamicschargedcompact}. In the next section, we will compute the waveform evolution using the cutoff $f_{\rm ISCO}$.

\subsection{Numerical waveforms}
We numerically integrate the angular-frequency evolution using a Runge–Kutta method and thereby obtain the time-domain GW phase. To plot the effect of the EM correction on the amplitude of the observable strain, we evaluate the time-domain waveform given in Eq. \eqref{h} for a representative equal-mass binary $m_1=m_2=10 M_\odot$ evolving from an initial GW frequency $f=50$ Hz to the corresponding cutoff $f=f_{\rm ISCO}$. 

\begin{figure}[htbp!]
    \centering
    \includegraphics[width=\linewidth]{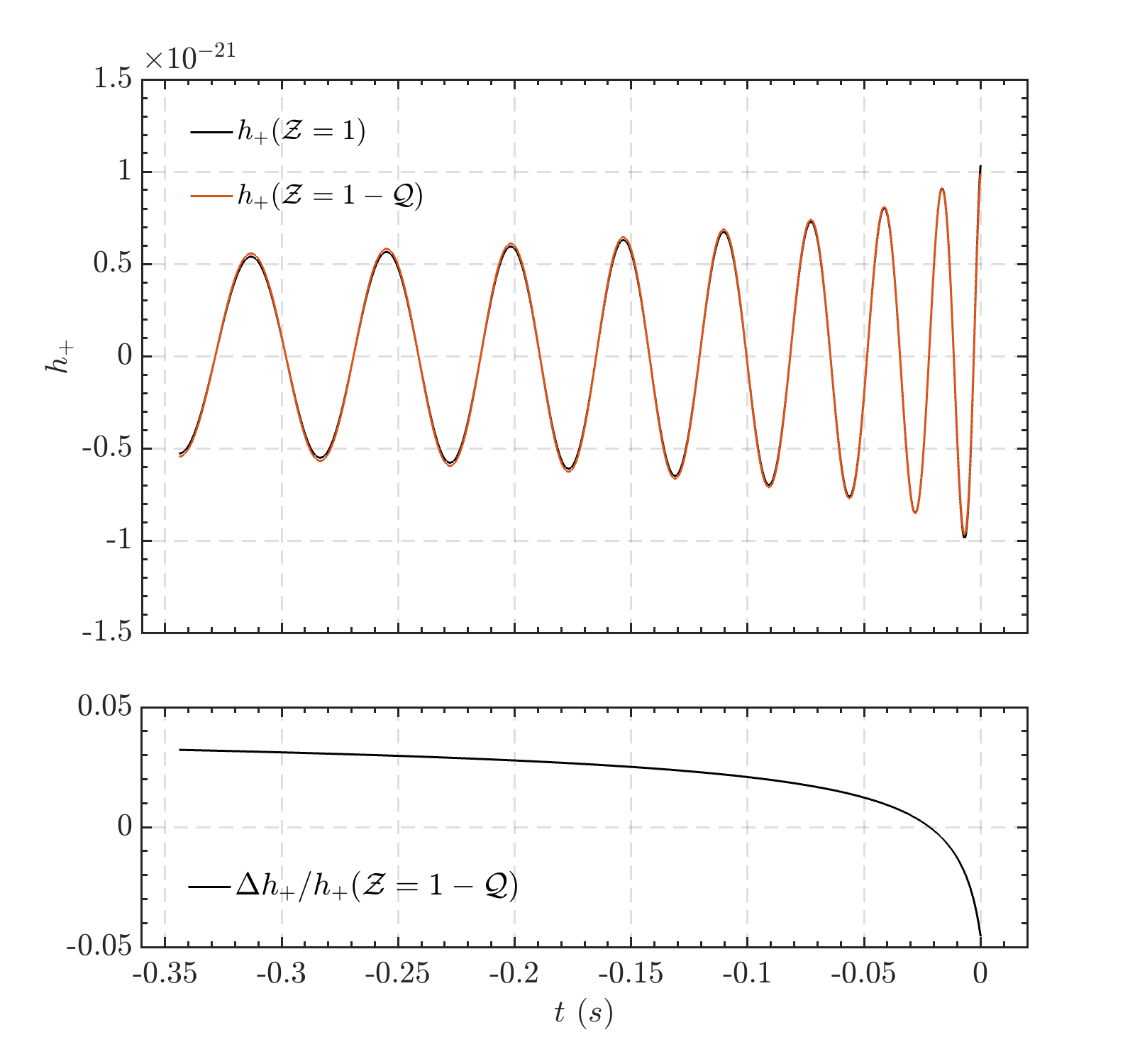}
    \caption{The EM corrections to the amplitude of plus polarization $h_+$ of the time-domain GW waveform for an equal-mass charged binary. 
In the upper panel, the black curve is obtained by suppressing the EM correction in the amplitude of waveform, $\mathcal{Z}=1$, 
whereas the red curve keeps the EM correction $\mathcal{Z}=1-\mathcal{Q}$. In the lower panel, we plot the fractional difference $\Delta h_+/h_+(\mathcal{Z}=1-\mathcal{Q})$, where $\Delta h_+=h_+(\mathcal{Z}=1-\mathcal{Q})-h_+(\mathcal{Z}=1)$.
The binary parameters are $m_1=m_2=10M_\odot$, $\kappa_1=-\kappa_2=0.3$, and $R=100$ Mpc, with an initial GW frequency $f_{0}=50~{\rm Hz}$. The time origin is chosen at ISCO.}
    \label{GW_Z}
\end{figure}

\begin{figure*}[htbp!]
    \centering
\includegraphics[width=0.49\linewidth]{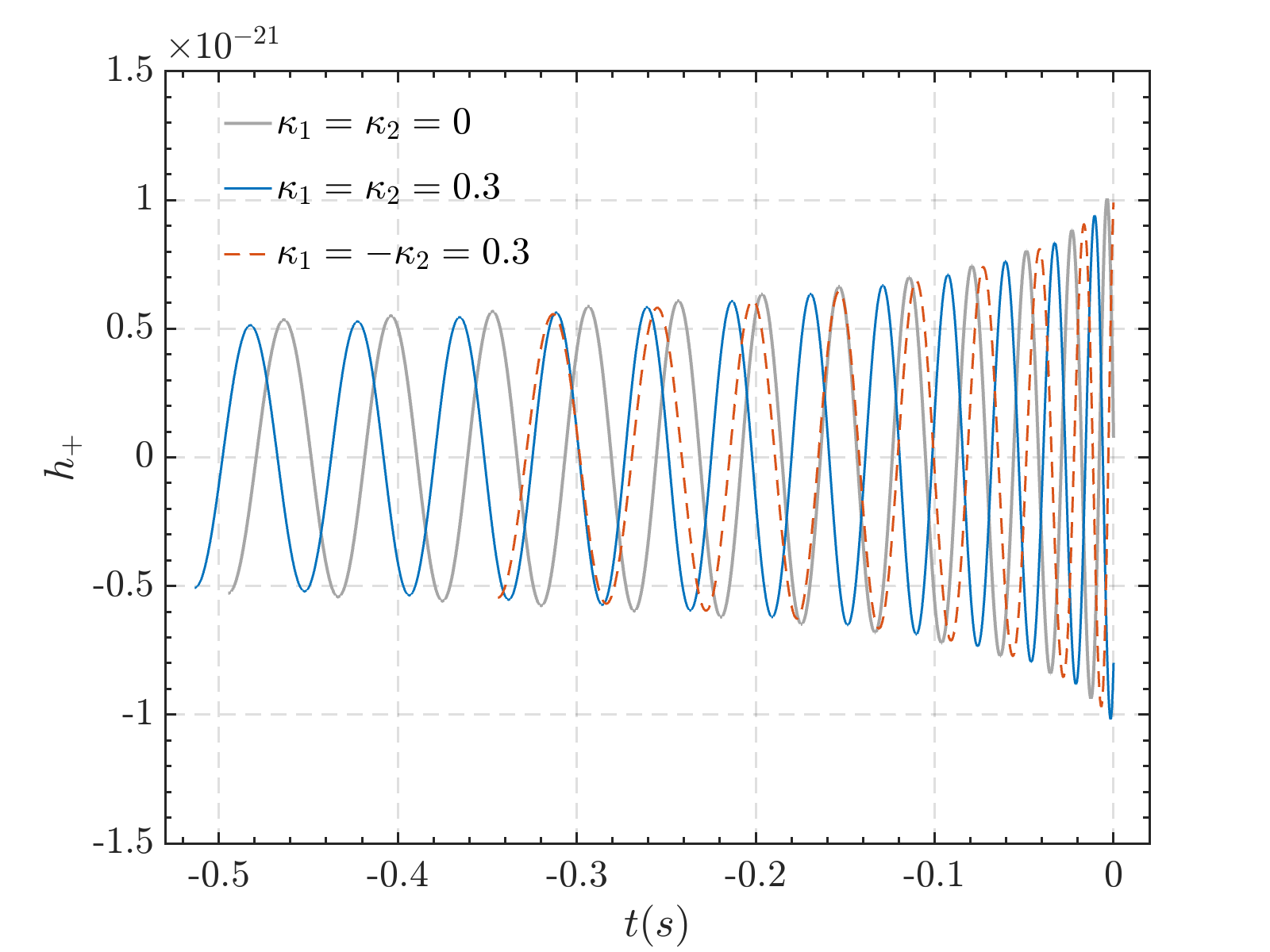}
    \includegraphics[width=0.49\linewidth]{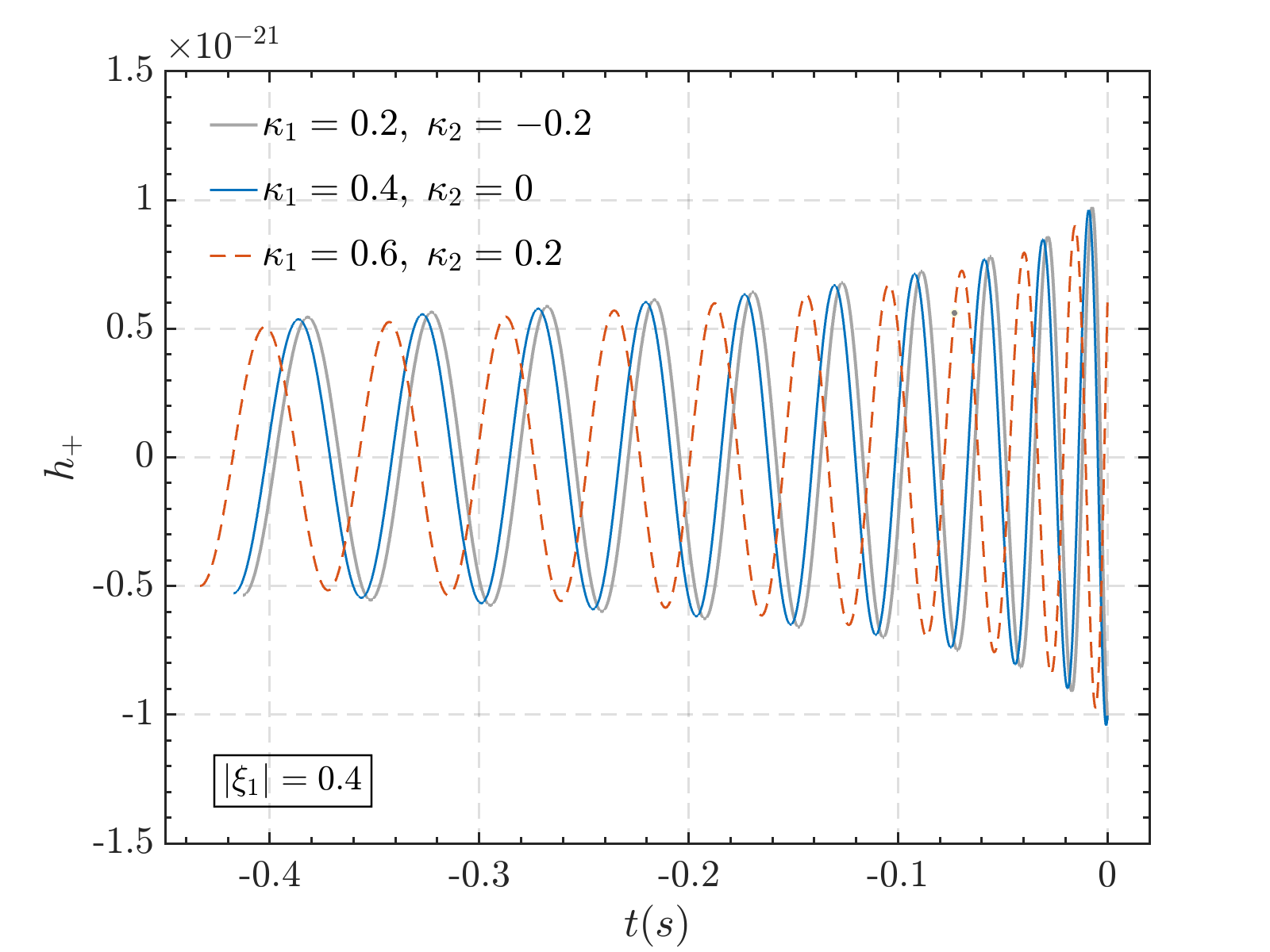}
    \caption{Plus polarization $h_+$ of the time-domain waveform for binaries with different charge-to-mass ratios. The left panel compares $(\kappa_1,~\kappa_2)=(0,0),~(0.3,~0.3),~\text{and} ~(0.3,~-0.3)$. The right panel compares $(0.2,~ -0.2),~ (0.4, 0),~ \text{and}~ (0.6,~ 0.2)$, all of which satisfy $|\xi_1|=|\kappa_1-\kappa_2|=0.4$. For all configurations, $m_1=m_2=10M_\odot$, initial frequency $f_{\text{GW}}=50\,{\text{ Hz}}$, $R=100$ Mpc, and the origin of time is chosen at the ISCO. Charges with the same sign and with opposite signs lead to different inspiral rates, thereby producing distinct phase accumulations and amplitude evolutions in the late inspiral waveform. }
    \label{GW_Q}
\end{figure*}

In Fig. \ref{GW_Z}, we compare the plus polarization obtained with $\mathcal{Z}=1$  with the result obtained by retaining the full factor $\mathcal{Z}=1-\mathcal{Q}$. 
The orbital phases $\phi$ are kept fixed in this comparison. 
The charged case exhibits a visible shift in the  amplitude. 
This behavior is consistent with the fact that the factor $\mathcal{Z}$ enters not only the affected Kepler relation, but also the 1PN correction to the mass quadrupole moment. The lower panel of Fig. \ref{GW_Z} shows a fractional difference of $h_+$ ranges from $-0.05$ to $0.03$ caused by the $\mathcal{Z}$ factor in the GW amplitude.

We plot the GW waveforms of binaries with different charge-to-mass ratios in Fig. \ref{GW_Q}. All binaries are equal-mass systems with $m_1=m_2=10 M_\odot$. In the left panel, compared to the neutral binary, the charged binaries accumulate a different orbital phase before reaching the ISCO. 
This dephasing results directly from the EM correction to the conservative dynamics and the additional EM contribution to the radiation flux. 
In particular, charges of opposite signs enhance the effective attraction and lead to faster coalescence, whereas charges of the same sign produce a phase evolution through the Coulomb correction. In the right panel, we plot the GWs generated by the binaries with the same charge-to-mass ratio difference $|\xi_1|=|\kappa_1-\kappa_2|=0.4$. For leading-order EM radiation only, EM contributions depend uniquely on $\xi_1^2$, yielding identical GW waveforms for equal $\xi_1^2$ \cite{Wang2021}. Higher-order EM radiation and electric multipole moments reduce the degeneracy between $\kappa_1$ and $\kappa_2$. In addition, differences in the ISCO frequencies among these charge configurations produce further waveform shifts.  

These numerical results show that EM effects leave a direct imprint on the inspiral waveform of charged compact binaries. 
For fixed masses and initial frequency, changing the charge-to-mass ratios affects both the rate of inspiral and the accumulated orbital phase. 
The comparison between the waveforms with and without the factor $\mathcal{Z}$ further indicates that the EM correction affects not only the conservative Keplerian relation, but also the waveform amplitude through the 1PN source quadrupole moment. 
These results motivate a more systematic construction of frequency-domain models including EM corrections.

\section{Frequency-domain waveforms}\label{Sec3}
\subsection{Stationary phase approximation}
We turn to the computation of the GW signal in the frequency domain, which provides the basis for constructing waveform models to be matched against GW observations.  Following the TaylorF2 construction \cite{PhysRevD.80.084043,PhysRevD.65.061501,PhysRevD.71.084008}, we compute the frequency-domain waveform using the stationary-phase approximation \cite{Maggiore:2007ulw}
\begin{align}
\Tilde{h}_+(f)=&\frac{1}{2}\mathcal{A}(t_f)\sqrt{\frac{\pi}{\ddot{\phi}(t_f)}}\left(\frac{1+c_\iota^2}{2}\right)e^{i\Psi_+},\\[2pt]
\Psi_+(f)=&2\pi ft_f-2\phi(t_f)-\frac{\pi}{4},\label{phit}
\end{align}
where $f=\omega/\pi$ is the GW frequency, and  $\dot\phi(t_f)=\pi f$. $t_f$ is the saddle point defined by solving for $t$. The GW amplitude is
\begin{align}\label{At}
    \mathcal{A}(t_f)=
\frac{4G^{5/3}\mathcal M_c^{5/3}}{c^4D}
\mathcal{Z}^{2/3}(\pi f)^{2/3}.
\end{align}

We define $t_c$ as the time at the ISCO and $D$ as the luminosity distance from the source to the detector. The chirp mass is $\mathcal{M}_c=m \eta^{3/5}$. Accounting for both electromagnetic and gravitational radiation, the balance equation gives \cite{damour1983gravitational,Damour1987,VanessaCdeAndrade_2001}
\begin{align}
    \frac{\mathrm{d}t_f}{ \mathrm{d}f}+\mathcal{F}(f)^{-1}\frac{d\mathcal{E}(f)}{df}=0.
\end{align}
Integrating, we obtain \cite{PhysRevD.80.084043}
\begin{align}
t(f)=&t_c+\int^{f_c}_{f}\mathcal{F}(f^\prime)^{-1}\frac{\mathrm{d}\mathcal{E}(f^\prime)}{\mathrm{d} f^\prime}\mathrm{d} f^\prime,
\end{align}
and the phase in Eq. \eqref{phit} can be rewritten as
\begin{equation}
    \Psi_+(f)=\Phi_c-\int^{f_c}_{f}2\pi t(f^\prime)\mathrm{d} f^\prime,\label{Psift}
\end{equation}
where $t_c$, $f_c$, and $\Phi_c$ are the coalescence time, GW frequency, and GW phase, respectively. The 1PN energy $\mathcal{E}(f)$ in the circular orbit is \cite{zhang2026postnewtoniandynamicschargedcompact}
\begin{align}
     \mathcal{E}(f)=&-\frac{\mu c^2}{2}\frac{\mathrm{x}}{\mathcal{Z}^{1/3}}\bigg\{\mathcal{Z} + \frac{\mathrm{x}}{\mathcal{Z}^{1/3}}\bigg[-\frac{3}{4}-\frac{\eta }{12}\notag\\
    &+\mathcal{Q} \left(-\frac{1}{2}+\frac{\zeta _0}{3}-\frac{7 \eta }{2}\right)+\mathcal{Q}^2 \left(\frac{7}{12}+\frac{17}{4}-\frac{\zeta_0}{3}\right)\notag\\
    &+\mathcal{Q}^3\left(-\frac{2 \eta }{3}-\frac{1}{6}\right)\bigg]\bigg\}+\mathcal{O}(c^{-4}),
\end{align}
where $\mathrm{x}\propto f^{2/3} $ is defined in Eq. \eqref{x}. The total flux is $\mathcal{F}=\mathcal{F}_{GW}+\mathcal{F}_{EM}$. Through 1PN order, the two contributions are
\begin{widetext}
\begin{align}
    \mathcal{F}_{GW}&=\frac{32}{5} \frac{c^5}{G}\eta^2  \mathcal{Z}^{4/3}x^{5}\bigg\{1+\frac{x}{\mathcal{Z}^{4/3}} \bigg[-8-4 \eta -2\zeta_0-4 \eta  \mathcal{Q}^2+ ( 10+8 \eta )\mathcal{Q}\notag\\
    &\qquad\qquad+\frac{2}{3}\bigg(6+\zeta_0+8 \eta +\frac{1}{2}(4\eta+1)\mathcal{Q}^2-(3+\eta)\mathcal{Q}\bigg)\mathcal{Z}+\left(\frac{97}{336}-\frac{17 \eta }{4}\right) \mathcal{Z}^{2}\bigg]\bigg\}+\mathcal{O}(c^{-9}),\label{FGW}\\
    \mathcal{F}_{EM}&=\frac{2}{3} \frac{c^5}{G}\eta^2 
    \mathcal{Z}^{2/3}x^4\bigg\{\xi_1^2+\frac{x}{\mathcal{Z}^{4/3}} \bigg[\bigg( \left(10  +8 \eta \right)\mathcal{Q}-4\eta -8-2\zeta_0 -4 \eta  \mathcal{Q}^2\bigg)\xi_1^2+\Big(\frac{12}{5}\zeta_2^2-\eta\xi_0\xi_1+\frac{4}{5}\xi_1\xi_3\Big)\mathcal{Z}^2\notag\\
    &\qquad\qquad+\frac{2}{3}\bigg(\Big(9+7 \eta +2\zeta_0+(1+4\eta)\mathcal{Q}^2+\left(-\frac{15}{2} -2 \eta  \right)\mathcal{Q}\Big) \xi_1^2-\frac{1}{3} \xi_1\xi_2\bigg)\mathcal{Z}\bigg]\notag\\
    &+\frac{24}{5}\frac{x^2}{\mathcal{Z}^{8/3}}\bigg[2\bigg(-2- \eta -  \eta \mathcal{Q}^2+ \zeta_0+ ( 2 \eta +1)\mathcal{Q}\bigg)\zeta_2^2+\bigg(-\frac{2}{21}\zeta_2\zeta_4-\frac{1219}{1120}(4 \eta -1)\xi_3^2\bigg)\mathcal{Z}^2\notag\\
    &\qquad\qquad+\bigg(\Big(2+\frac{8}{3}\eta+\frac{1}{3}\zeta_0+\frac{1}{6} (1+4\eta)\mathcal{Q}^2+\frac{1}{3}\left(3+ \eta\right)\mathcal{Q}\Big)\zeta_2^2- \zeta_2\zeta_3\bigg)\mathcal{Z}\bigg]\bigg\}+\mathcal{O}(c^{-9}),\label{FEM}
\end{align}
\end{widetext}
We define the following dimensionless EM combinations of the charge-to-mass ratios and mass fractions:
\begin{gather}
\xi_0\equiv(\kappa_1\chi_1+\kappa_2\chi_2)(\chi_1-\chi_2),\notag\\
\xi_1\equiv\kappa_1-\kappa_2,\quad\xi_2\equiv\kappa_1\chi_2-\kappa_2\chi_1,\notag\\
\xi_3\equiv\kappa_1\chi_2^2-\kappa_2\chi_1^2,\notag\\
\zeta_2\equiv\kappa_1\chi_2+\kappa_2\chi_1,\quad
\zeta_3\equiv\kappa_1\chi_2^2+\kappa_2\chi_1^2,\notag\\
\zeta_4\equiv\kappa_1\chi_2^3+\kappa_2\chi_1^3,
\end{gather}
where $\chi_1=m_1/m$ and $\chi_2=m_2/m$ are the fractions of the mass of the components. All $\xi_i$ and $\zeta_i$ are linear in $\kappa_i$, except $\zeta_0$ is quadric in $\kappa_i$. 
In the frequency domain, the two GW polarizations take the form
\begin{align}
 \Tilde{h}_+(f)=&\mathcal{A}(f)\left(\frac{1+c_\iota^2}{2}\right)e^{i\Psi_+(f)},\\
 \Tilde{h}_\times(f)=&\mathcal{A}(f)c_\iota e^{i\Psi_\times(f)}.
\end{align}
Next, we derive analytic expressions for $\Psi_+(f)$ using the following two cases.  {Note that since the leading electric-dipole radiation appears at $-1$ PN order, the electric-dipole contribution derived here is accurate up to 0 PN order. A complete 1 PN-accurate treatment would require the next-to-next-to-leading order  electric-dipole radiation and the 2PN conservative dynamics, which are beyond the scope of this paper.}

\subsection{Gravitational-quadrupole dominated waveforms}

To obtain an analytic PN phase, we expand $\Psi(f)$ in the powers of ${\rm{x}}\propto f^{2/3}$. The leading EM contribution enters at -1PN order. In order to expand $\Psi(f)$ correctly, we need to treat $\kappa_1^2\thicksim \kappa_2^2\thicksim \mathcal{Q}\ll 1$. We retain terms through $\kappa^2$ as the leading-order EM contribution and neglect $\mathcal{O}(\kappa^3)$ EM terms. This expansion is valid under the condition $\mathrm{x}\gg \rm{x}_E$.  {Setting the gravitational-quadrupole effect equal to the electric-dipole effect in Eq. \eqref{Domega}, we obtain}
\begin{equation}
    \rm{x}_E\equiv 5/48\xi_1^2\mathcal{Z}^{4/3}.
    \end{equation} 
The corresponding GW frequency is
\begin{equation}\label{fE}
    f_{\rm{E}} =7.41 \text{Hz} \left(\frac{m}{20 M_\odot}\right)^{-1}\left(\frac{|\xi_1|}{0.4}\right)^3\left(\frac{\mathcal{Z}}{1.04}\right)^2.
\end{equation}

The amplitude of the GWs is obtained from Eq.~\eqref{At}
\begin{align}\label{AH}
    \mathcal{A}^G(f)=&\frac{1}{D}\sqrt{\frac{5}{24}}\pi^{-2/3} c^{-3/2}(G \mathcal{M}_c)^{5/6} f ^{-7/6}\notag\\
    &-\frac{1}{D}\frac{5}{192}\sqrt{\frac{5}{6}}\pi^{-4/3} \xi_1^2 \eta ^{2/5} c^{1/2}(G\mathcal{ M}_c)^{1/6}f ^{-11/6}\notag\\[6pt]
    &+\mathcal{O}(\kappa^3),
 \end{align}
where the  {superscript $G$ denotes the} gravitational-quadrupole dominated expansion. The first term is the leading gravitational contribution, whereas the second term is the leading EM contribution. Time $\tau^G(f)=t_c-t^G(f)$ is
\begin{widetext}
\begin{align}\label{tH}
    \tau^G(f)=&\frac{5}{256}\frac{G }{c^3}\frac{m}{\mathrm{x}^4\eta}\bigg[1+\frac{1}{252}\mathrm{x}(743+924\eta)\bigg]\notag\\
    &+\frac{5 }{3072}\frac{G }{c^3}\frac{m}{\mathrm{x}^5 \eta }\bigg\{- \xi _1^2+\mathrm{x} \bigg[-\xi _1^2\left(\frac{95 }{12}\eta +\frac{3295 }{672}\right) -3 \zeta _2^2+\frac{5}{4}  \xi _0 \xi _1+\frac{5 }{2}\xi _1 \xi _2-\xi _1 \xi _3+8\mathcal{Q}\bigg]\notag\\
    &+\mathrm{x}^2 \bigg[\frac{64 }{3}\zeta _0 -28 \eta\zeta _2^2  + 8\zeta _2  \zeta _3+\frac{16}{21} \zeta _2\zeta _4-\frac{323 }{42}\zeta _2^2-  \frac{5}{48} \xi _1^2\left(\frac{4079}{9}\eta^2 +\frac{10657 }{14}\eta+\frac{690257 }{2352}\right)+\frac{1219}{140}\xi _3^2\left(4\eta-1\right)\notag\\
    &+ \frac{5}{9}\xi _0 \xi _1\left(17 \eta +\frac{995}{56} \right)+  \frac{5}{9}\xi _1 \xi _2\left(34 \eta+\frac{995 }{28}\right)-  \frac{1}{9}\xi _1 \xi _3\left(68 \eta+\frac{995 }{14}\right)-\mathcal{Q}192 \eta \bigg]\bigg\}+\mathcal{O}(\kappa^3,\mathrm{x}^{-2}).
\end{align}
Integrating according to Eq. \eqref{Psift}, we obtain the GW phase
\begin{align}\label{Psi}
    \Psi^G_+(f)=&2\pi f t_c-\Phi_c-\frac{\pi}{4}+\frac{3}{128  \eta {\rm x}^{5/2}}\left\{1+ {\rm x}\frac{5}{9}\left(\frac{743}{84}+11 \eta \right) \right\}\notag\\
    &+\frac{5}{3584\eta {\rm x}^{7/2}}\bigg\{-\xi_1^2+ {\rm x}\bigg[-\frac{21 }{5}\zeta _2^2+\left(-\frac{133 \eta }{12}-\frac{659}{96}\right) \xi _1^2+ \frac{7 }{4} \xi _0\xi _1+\frac{7 }{2}\xi _1\xi _2-\frac{7 }{5}\xi _1\xi _3+\frac{56}{5}  \mathcal{Q} \bigg]\notag\\
    &+ {\rm x}^2\frac{1}{3}\bigg[56 \zeta _2 \zeta _3-196 \zeta _2^2 \eta +\frac{16}{3}\zeta _2 \zeta _4-\frac{323}{6}\zeta _2^2-\frac{5}{96}\xi _1^2\left(\frac{57106}{9} \eta ^2+10657 \eta +\frac{690257}{168}\right)+\frac{5}{9}\xi _1\xi _0 \left(119 \eta +\frac{995}{8}\right) \notag\\
    &+\frac{5}{9}\xi _1\xi _2\left(238 \eta+\frac{995}{4}\right)-  \frac{1}{9}\xi _1\xi _3\left(476 \eta+\frac{995}{2}\right)+\frac{1219}{20} \xi _3^2\left( 4\eta -1\right)+\frac{448}{3} \zeta _0 -1344\mathcal{Q} \eta\bigg]\bigg\}\notag\\[6pt]
    &+\mathcal{O}(\kappa^3,\mathrm{x}^{-1/2}),
\end{align}
\end{widetext}
and $\Psi_+=\Psi_\times+\pi/2$. Here we expand $\mathcal{Z}=1-\mathcal{Q}$. Retaining $\mathcal{Z}$ in $\Psi^G(f)$, we obtain the same  {-1PN EM contribution} as \cite{Wang2021,Cardoso_2016,Christiansen_2021}.
The leading-order gravitational quadrupole radiation dominates the phase $\Psi_+^G(f)$ and is proportional to $\rm{x}^{-5/2}$, whereas the leading-order electromagnetic term is proportional to $\rm{x}^{-7/2}$ and produces a contribution much weaker than that of the leading-order gravitational radiation with $\rm{x}\gg \rm{x}_E$.
These results are valid in the gravitational-quadrupole dominated regime, but break down in the electric-dipole dominated regime. In the neutral limit $\kappa_1=\kappa_2=0$,  the quantities $\mathcal{A}^G(f),~\tau^G(f)$ and $\Psi^G(f)$ reduce to the standard PN expressions.

In agreement with the calculation of the time-domain, the contribution of $-1$PN EM to $\Psi^G_+(f)$ depends only on charge-to-mass ratio difference $\xi_1^2=(\kappa_1-\kappa_2)^2$. The 0PN and 1PN EM corrections introduce additional charge combinations ratio $\xi_i$ and $\zeta_i$ that can reduce the degeneracy between  $\kappa_1$ and $\kappa_2$.

Note that the 1PN EM contribution is incomplete.

\subsection{Electric-dipole dominated waveforms}
We derive a complementary expansion for the electric-dipole dominated regime $\mathrm{x}\ll \rm{x}_E$. The GW amplitude is 
\begin{equation}\label{AL}
    \mathcal{A}^E(f)=\frac{1}{D}\frac{1}{c^{5/2}}\sqrt{2} \pi^{-1/3} {\eta }^{-1/5}(G \mathcal{M}_c)^{7/6}\frac{\mathcal{Z}^{2/3}}{|\xi_1|}f^{-5/6}.
\end{equation}
The time $\tau^E(f)=t_c-t^E(f)$ is

\begin{widetext}
\begin{align}\label{tL}
    \tau^E(f)=&\frac{G }{c^3}\frac{m}{4 \mathrm{x}^3\eta}\bigg\{\xi _1^{-2}+\frac{\mathrm{x}}{\mathcal{Z}^{4/3}}\xi _1^{-4}\bigg[-\frac{72}{5}-2\zeta _0 \xi _1^2-\frac{18 \zeta _2^2}{5}+\left(\frac{3}{4}-\frac{5 \eta  }{4}\right)\xi _1^2+\frac{3 \xi _0 \xi _1}{2}+3 \xi _1 \xi _2-\frac{6 \xi _1 \xi _3}{5}\notag\\
    &+\mathcal{Q} \left(6 \zeta _0 \xi _1^2+\frac{36 \zeta _2^2}{5}-\frac{27}{2}\eta\xi _1^2-3 \xi _0 \xi _1-3 \xi _1 \xi _2+\frac{12 \xi _1 \xi _3}{5}+\frac{144}{5}\right)\notag\\
    &+\mathcal{Q}^2 \left(-\zeta _0 \xi _1^2-\frac{18}{5} \zeta _2^2+\frac{41 }{4}\eta\xi _1^2-\frac{17}{4}\xi _1^2+\frac{3 \xi _0 \xi _1}{2}-\frac{6 \xi _1 \xi _3}{5}-\frac{72}{5}\right)+\mathcal{Q}^3 \left(\frac{1}{2}+2 \eta \right)\xi _1^2\bigg]\bigg\}+\mathcal{O}(x^{-2}).
\end{align}
The GW phase is
\begin{align}\label{PsiL}
    \Psi^E_+(f)=&2\pi f t_c-\Phi_c-\frac{\pi}{4}+\frac{1}{ 2\eta \mathrm{x}^{3/2}}\bigg\{\xi_1^{-2}+\frac{3}{4}\frac{\mathrm{x} }{\mathcal{Z}^{4/3}}\xi_1^{-4}\bigg[-\frac{36}{5} \zeta _2^2-\frac{144}{5}-\xi _1^2 \left(4 \zeta _0+\frac{5 \eta }{2}-\frac{3}{2}\right)+\left(3 \xi _0+6 \xi _2-\frac{12 \xi _3}{5}\right) \xi _1\notag\\
    &+\mathcal{Q} \left(\xi _1^2 \left(12 \zeta _0-27 \eta \right)+\frac{72 \zeta _2^2}{5}+\left(-6 \xi _0-6 \xi _2+\frac{24 \xi _3}{5}\right) \xi _1+\frac{288}{5}\right)\notag\\
    &+\mathcal{Q}^2 \left(\xi _1^2 \left(-2 \zeta _0+\frac{51 \eta }{2}-\frac{17}{2}\right)-\frac{1}{5} 36 \zeta _2^2+\left(3 \xi _0-\frac{12 \xi _3}{5}\right) \xi _1-\frac{144}{5}\right)+\mathcal{Q}^3(4 \eta +1) \xi _1^2 \bigg]\bigg\}+\mathcal{O}(x^{1/2}),
\end{align}
\end{widetext}
where the leading order of the GW phase becomes $\Psi^E(f)\propto\mathrm{x}^{-3/2}$.  {To obtain the above result, we have ignored the 1PN terms in Eqs. \eqref{FGW} and \eqref{FEM}.}
 The quantities $\mathcal{A}^E(f),~\tau^E(f)$ and $\Psi^E(f)$ diverge as $\xi_1^{-1}\rightarrow \infty$, reflecting that the neutral limit lies outside the domain of validity of the large-charge expansion. 
 The two  phases $\Psi^E$ and $\Psi^G$ do not match continuously and therefore do not cover the full frequency range. Constructing a uniformly valid frequency-domain waveform is left for future work.

\subsection{Waveform with different charges}
\begin{figure}[htbp!]
    \centering
    \includegraphics[width=\linewidth]{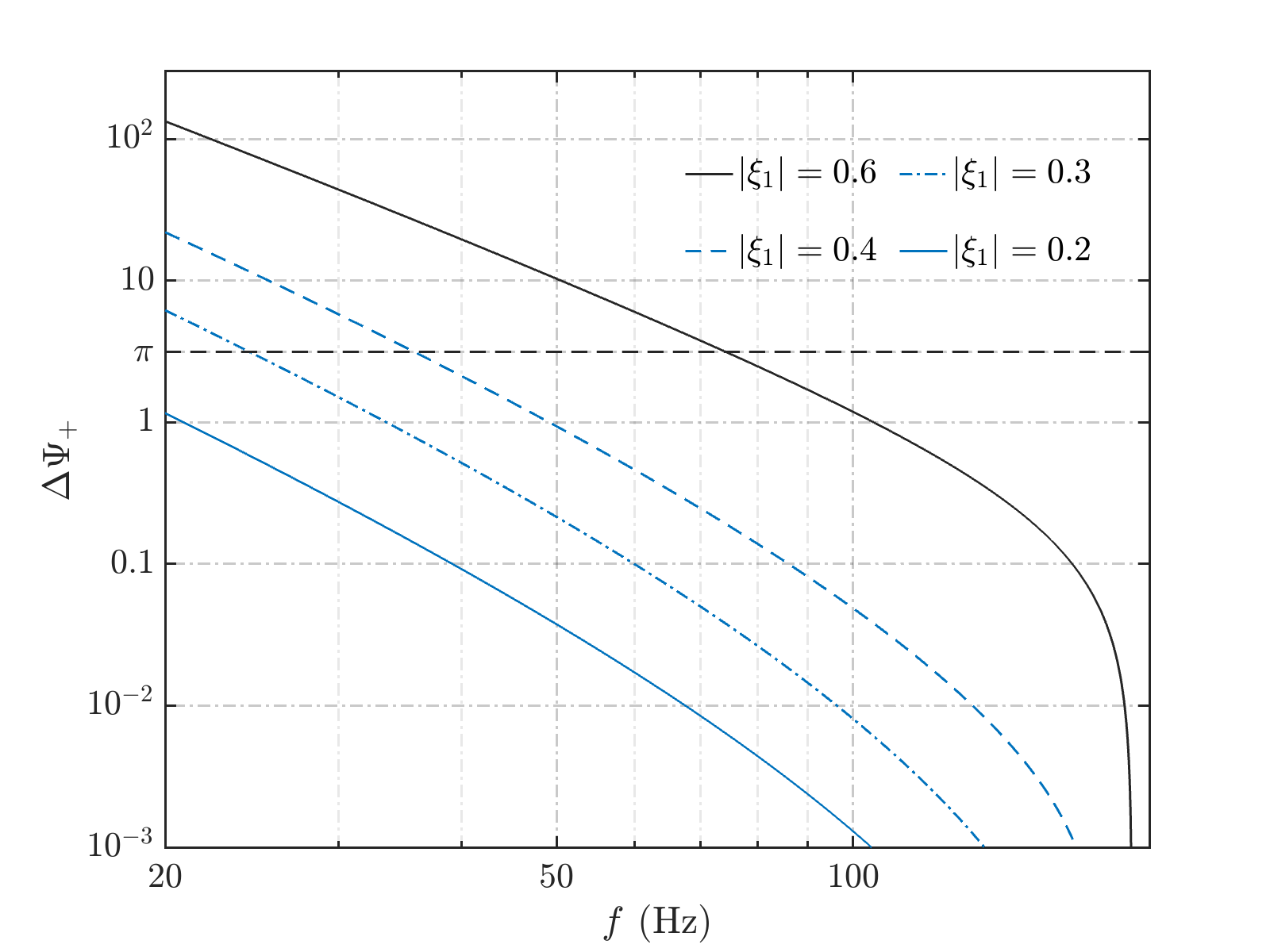}
    \caption{The error and valid frequency range of the GW phase $\Psi^G_+$.
The phase difference is defined as $\Delta \Psi_+=\Psi_+^T-\Psi_+^G$, where $\Psi_+^G$ denotes the phase obtained from the gravitational-quadrupole dominated expansion, and $\Psi_+^T$ denotes the phase computed via  numerical integration of Eq. \eqref{phit}. Because $\Psi_+^T$ is computed without an asymptotic expansion in the charge parameters, it serves as the reference phase. For equal-mass binaries with masses $m_1=m_2=10 ~M_\odot$, we plot the phase error across frequencies spanning $f_0=20$ Hz up to the ISCO frequency $f_{\rm ISCO}$  with charge-to-mass ratios $(\kappa_1,\kappa_2)=(0.3,-0.3),~(0.2,-0.2),~(0.2,-0.1),~\text{and}~(0.1,-0.1)$. We set $t_c=0$ and choose $\Phi_c$ to obtain $\Psi^G_+(f_{\rm ISCO})=\Psi^T_+(f_{\rm ISCO})=0$.}
    \label{Delta_Psi}
\end{figure}

Gravitational-wave events of GWTC-5.0 lie at frequencies above 20 Hz, which motivates an analysis of the valid regime for the gravitational-quadrupole dominated expansion. In Fig.~\ref{Delta_Psi}, we plot the GW phase difference $\Delta\Psi$. Here, $\Psi^T$ denotes the phase obtained from time-domain numerical integration without the asymptotic expansion in charge and acts as a reference to identify the valid frequency band of $\Psi^G$. The figure presents phase differences for four distinct charge-to-mass-ratios. The phase $\Psi^G$ becomes entirely unreliable once $|\Delta \Psi|>\pi$. Consequently, valid Bayesian inference of GWTC-5.0 events using $\Psi^G$ is restricted to $|\xi_1|<0.3$, or equivalently $\xi_1^2<0.09$. For an equal-mass binary with masses $m_1=m_2=10 M_\odot$ at the frequency $f_0=20$ Hz, we obtain $(\xi^E_1)^2=48/5\mathcal{Z}^{-4/3}( G m \pi f_0/c^3)^{2/3}\thicksim 0.33$ and $\xi_1^2\ll (\xi^E_1)^2$.

\begin{figure*}[ht!]
    \centering
    \includegraphics[width=\linewidth]{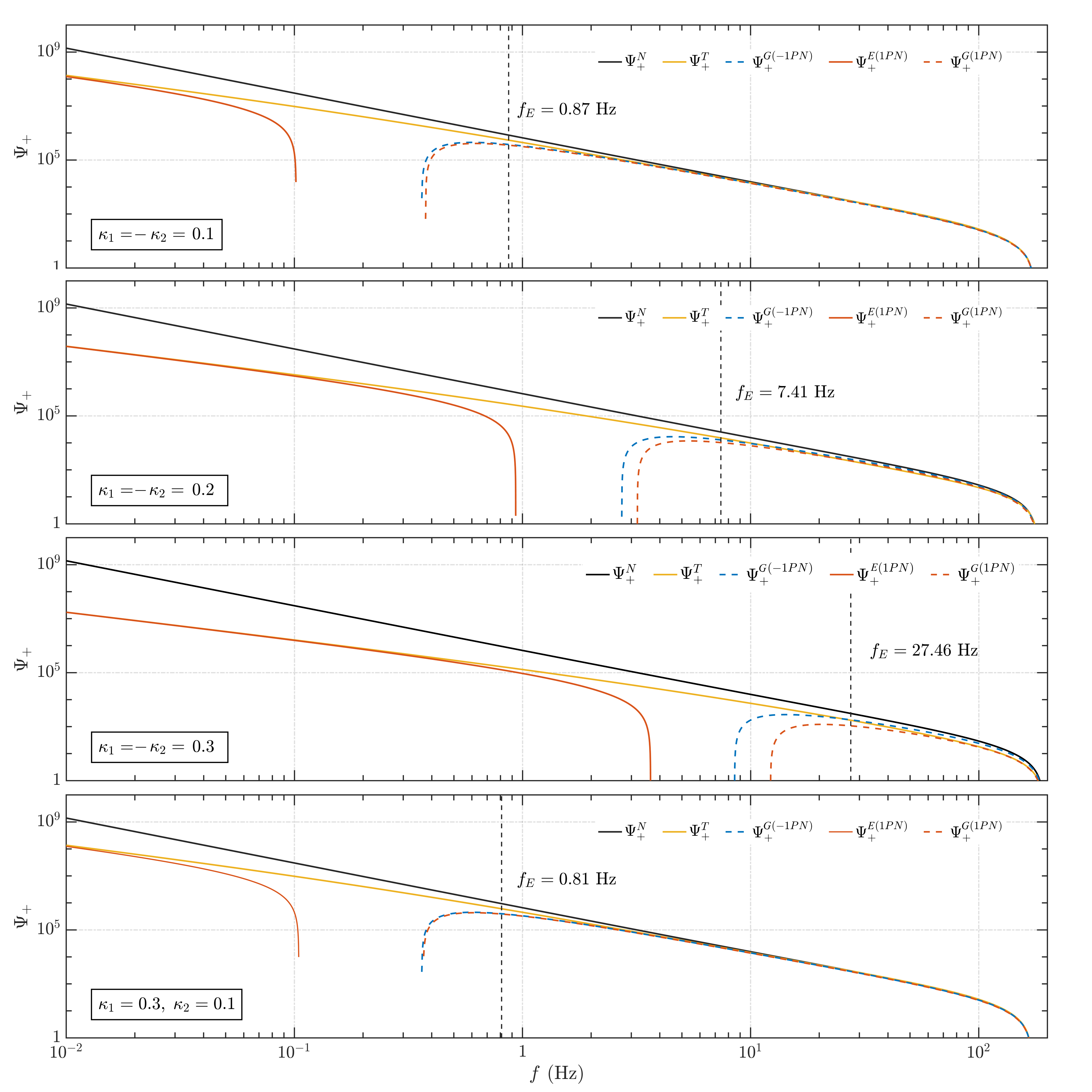}
    \caption{Frequency-domain phase $\Psi_+(f)$  from $f_0=10\,{\rm mHz}$ to the ISCO frequency $f_{\rm ISCO}$ for charged binaries with equal mass $m_1=m_2=10~ M_\odot$. 
$\Psi^N_+$ denotes the GW phase of the neutral binary, $\Psi^{G(-1PN)}_+$ and $\Psi^{G(1PN)}_+$ denote the GW phases in the gravitational-quadrupole dominated expansion $f\gg f_E$ with the -1PN and  the up to 1PN EM contributions in Eq. \eqref{Psi} respectively. $\Psi^{E(1PN)}_+$ denotes the GW phase in the electric-dipole dominated expansion $f\ll f_E$ with the 1PN EM contributions in Eq. \eqref{PsiL}, and $\Psi^T_+$ denotes the GW phase  via  numerical integration of Eq. \eqref{phit}.
The four panels show the phases for the binaries with charge-to-mass ratios $(\kappa_1,\kappa_2)=(0.1,-0.1),~(0.2,-0.2),~(0.3,-0.3),~\text{and}~(0.3,0.1)$. We set $t_c=0$ and choose $\Phi_c$ to obtain $\Psi_+(f_{\rm ISCO})=0$.}
    \label{EMPN}
\end{figure*}

\begin{figure*}[htbp!]
    \centering
\includegraphics[width=0.49\linewidth]{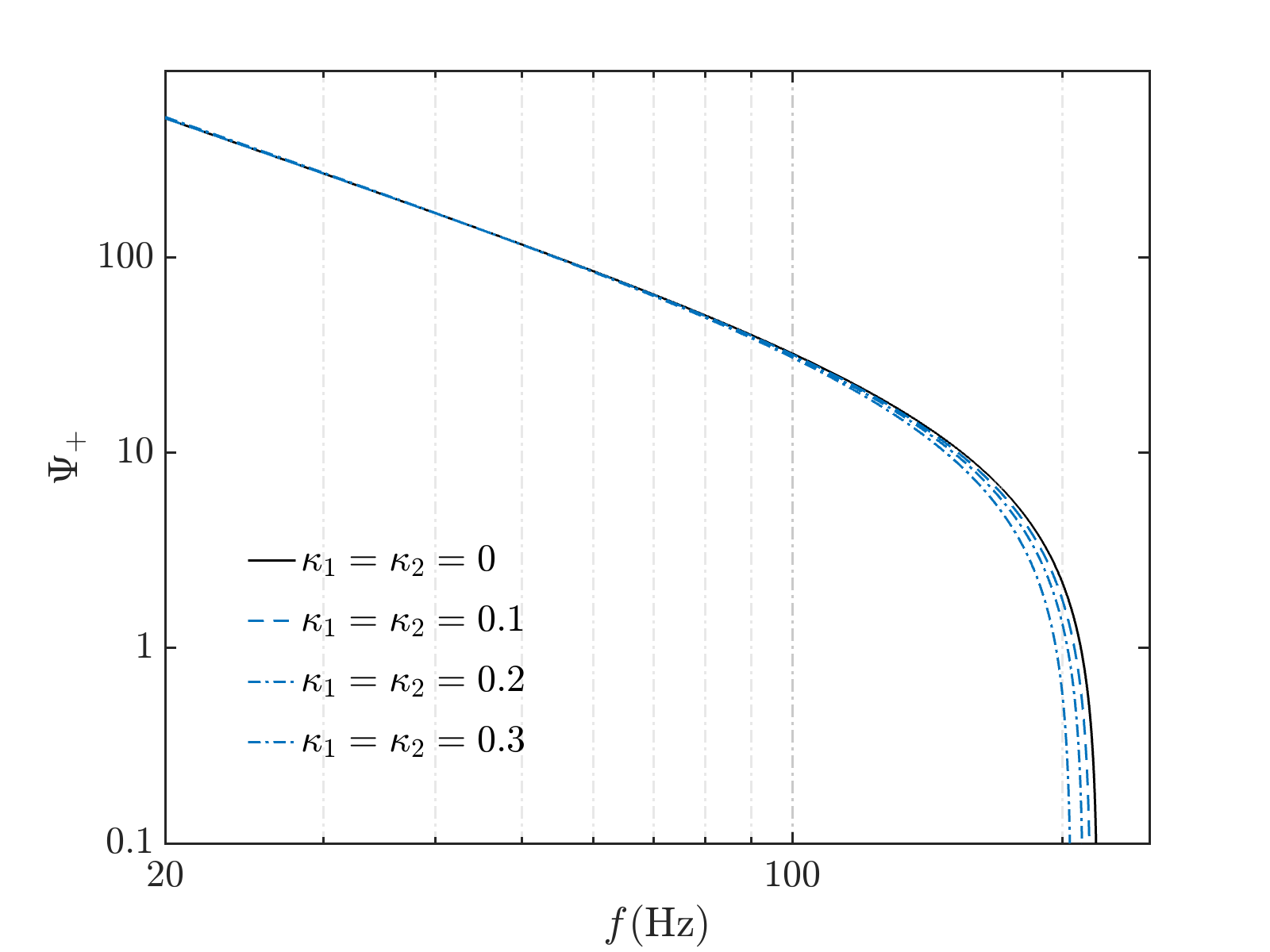}
    \includegraphics[width=0.49\linewidth]{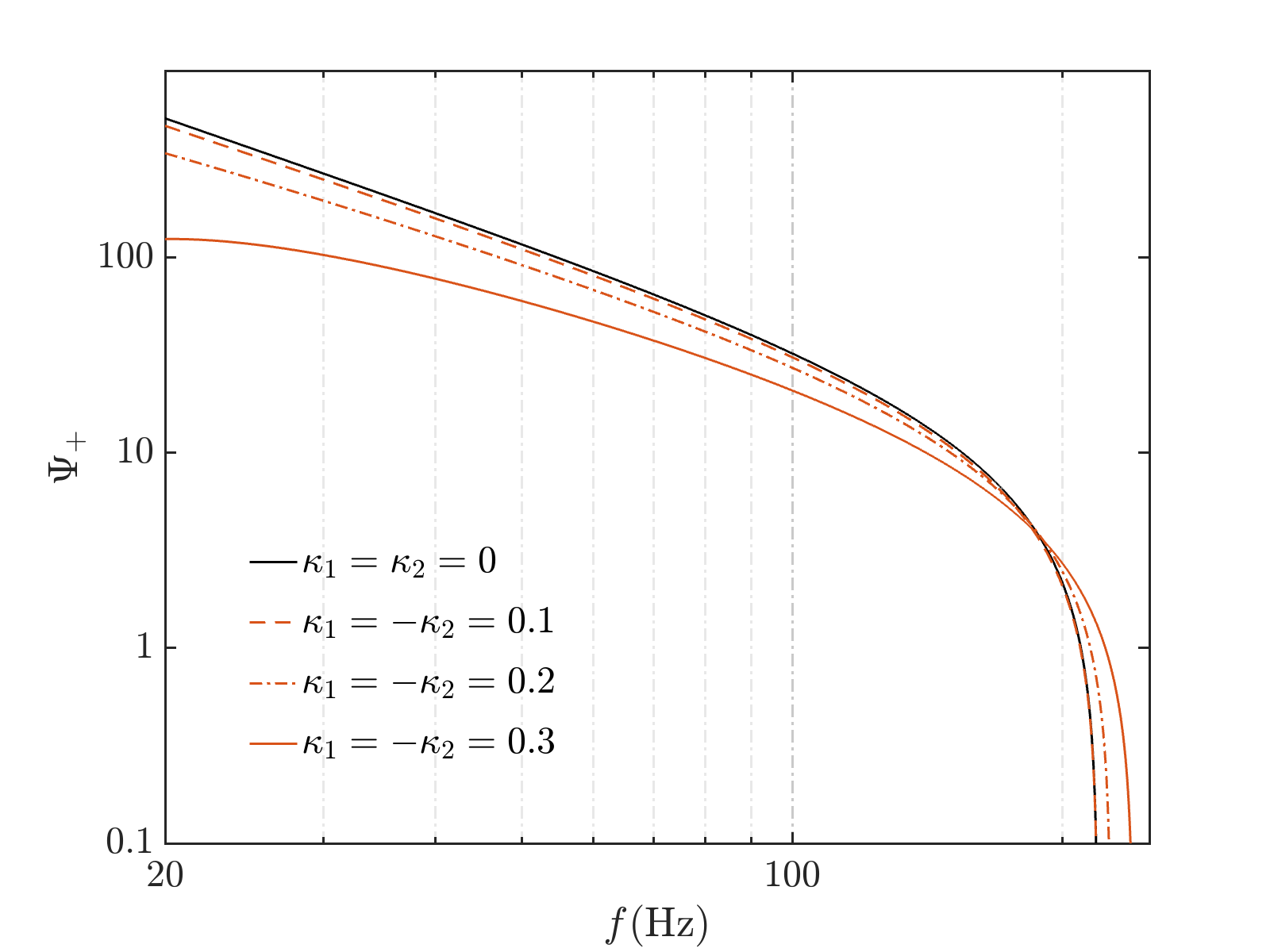}
    \caption{Frequency-domain phase $\Psi^G_+(f)$ for equal-mass charged binaries in the gravitational-quadrupole dominated regime with different charge-to-mass ratios. 
The left panel shows binaries with charges of the same sign, $\kappa_1=\kappa_2$, while the right panel shows binaries with charges of opposite signs, $\kappa_1=-\kappa_2$. 
 Solid black curves denote the neutral binaries. 
 For the same $\kappa_i$, opposite-sign binaries produce a larger phase shift than same-sign binaries.
The endpoint of each curve is set by the corresponding ISCO frequency. 
As $|\kappa_i|$ increases, $f_{\rm ISCO}$ decreases for charges of the same sign, whereas it increases for charges of opposite signs.}
    \label{Psi_SP}
\end{figure*}

Fig.~\ref{EMPN} compares the frequency-domain phases in Eqs.~\eqref{Psi} and \eqref{PsiL} for charged binaries with representative charge-to-mass ratios between $0.1$ and $0.3$.
The phase is evaluated in the interval $10\,{\rm mHz}\leq f\leq f_{\rm ISCO}$. 
With the chosen charge-to-mass ratio of the binaries, only the GW phase  $\Psi^E$ is valid at low frequencies, as plotted by the solid red line. In contrast, only the GW phase $\Psi_S$ is valid at high frequencies, corresponding to the red dashed and blue solid curves. The respective expansion regimes impose the constraints $f\ll f_E$ and $f\gg f_E$, where $f_E=({\rm x}_E)^{3/2}c^3/(Gm\pi)$ in Eq. \eqref{fE} is marked by the vertical dashed black lines, such that both gravitational-quadrupole and electric-dipole dominated expansions break down in the transition frequency band and fail to cover the full frequency domain. We compare $\Psi^G_+$ and $\Psi^E_+$ with $\Psi^T_+$, which is plotted by the yellow curves. $\Psi^T_+$ is obtained by numerically integrating Eqs.~\eqref{Domega} and \eqref{Psift} in the time domain and is valid across the full frequency range considered.
The transition frequency band increases as the difference in the charge-to-mass ratio  $|\xi_1|$ increases. Relative to the leading-order treatment used in Refs. \cite{Wang2021,sym15020537}, (blue dotted curves in Fig.~\ref{EMPN}), our high-frequency phase includes higher-PN EM corrections, while the large-charge expansion extends the analytic treatment to the electric-dipole dominated regime for fixed charge-to-mass ratios.

We compare the dependence of the frequency-domain phase on the relative signs of the two charges. 
Fig.~\ref{Psi_SP} shows $\Psi^G_+(f)$ for equal-mass binaries, with the same-sign case  $\kappa_1=\kappa_2$ in the left panel and the opposite-sign case  $\kappa_1=-\kappa_2$ in the right panel.
In both panels, the neutral binary is shown as a reference. 

For binaries with charges of the same sign, Coulomb repulsion weakens the effective attraction and produces a moderate dephasing with respect to the neutral waveform. 
The corresponding ISCO frequency decreases as the charge-to-mass ratio increases.
As the magnitude of equal charges  of the same sign increases, the impact of the EM sector becomes more pronounced. This behavior is absent in leading-dipole analyses because the dipole contribution vanishes when $\xi_1=\kappa_1-\kappa_2=0$ for binaries with equal charge-to-mass ratios $\kappa_1=\kappa_2$. Therefore, in the left panel, the EM contribution to the inspiral is influenced by higher-order multipolar radiation, in addition to the GW radiation.

For binaries with charges of opposite signs, Coulomb interaction is attractive and the phase correction becomes much more pronounced. 
In this case the ISCO frequency increases as the charge-to-mass ratio increases, and the accumulated phase differs substantially from the neutral result over the whole plotted frequency interval. 
This comparison shows that the sign of the charge product is a key parameter controlling both the ISCO of the inspiral and the contributions of the electromagnetic correction to the frequency-domain phase.

\section{Bayesian Inference Results}\label{Sec4}
\begin{table*}[htbp!]
\small
  \centering
  \caption{Constraints on charge-to-mass ratios $\kappa_1,~\kappa_2$ and $\mathcal{Q}_1$ and masses $m_1,~m_2$ and $\mathcal{M}_c$ obtained from three selected O4b BBH events, 68\%
credible interval is shown.}
    \renewcommand\arraystretch{2}
    \setlength{\tabcolsep}{30pt}
    \begin{tabular}{ccccc}
    \hline
     Event name&GW240925  & GW241102 & GW250119 \\
    \hline 
     $\kappa_1$ & $-0.069^{+0.087}_{-0.071}$ &  $-0.078^{+0.084}_{-0.065}$ & $0.050^{+0.166}_{-0.120}$ \\
    $\kappa_2$ & $0.123^{+0.086}_{-0.072}$ &  $0.106^{+0.081}_{-0.063}$ & $0.113^{+0.164}_{-0.081}$ \\
       $\mathcal{Q}$ & $-0.006^{+0.010}_{-0.004}$ &  $-0.006^{+0.007}_{-0.003}$ & $0.005^{+0.037}_{-0.008}$ \\
    $m_1~(M_\odot)$ & $8.583^{+0.806}_{-0.511}$ & $9.996^{+0.722}_{-0.454}$ & $12.416^{+1.199}_{-0.734}$ \\
    $m_2~(M_\odot)$ & $7.282^{+0.410}_{-0.605}$ & $8.819^{+0.417}_{-0.588}$ & $10.307^{+0.756}_{-1.287}$ \\
    $\mathcal{M}_c~(M_\odot)$ & $6.882^{+0.136}_{-0.167}$ & $8.140^{+0.188}_{-0.114}$ & $9.821^{+0.109}_{-0.233}$ \\
    \hline
    \end{tabular}
  \label{constrain}
\end{table*}

 
Electric charges in BBHs affect the orbital dynamics, making GW observations a direct probe of possible charge effects. In this section, we carry out Bayesian inference \cite{PhysRevD.80.122003,PhysRevD.91.042003,Thrane_Talbot_2019,PhysRevD.49.2658} with B{\footnotesize ILBY} \cite{Ashton_2019}, using {\ttfamily dynesty} \cite{10.1093/mnras/staa278} as the nested-sampling algorithm. We analyze three selected low-mass BBH events from the fourth observation run of the LIGO–Virgo–KAGRA Collaboration \cite{theligoscientificcollaboration2026opendataligovirgo,theligoscientificcollaboration2026gwtc50observationssecondfourth}. We focus on GW$240925\_005809$, GW$241102\_ 124058$, and GW$250119\_190238$ in  GWTC-5 \cite{theligoscientificcollaboration2026gwtc50observationssecondfourth, gzrj-mwv3,Abac_2025}, which contain a relatively large number of inspiral cycles in the detector sensitive band and have relatively high signal-to-noise ratios. These events are therefore well suited for constraining charge-induced dipole radiation effects within the PN waveform framework adopted in this work.

In this work, Bayesian inference is performed with a hybrid waveform model
\begin{align}
 \hat{h}_+(f)=&\mathcal{A}^G(f)\left(\frac{1+c_\iota^2}{2}\right)e^{i\hat{\Psi}^G_+(f)}.
\end{align}
We take the amplitude to be $\mathcal{A}^G$ from Eq.~\eqref{AH}. To construct the phase $\hat{\Psi}^G$, we add the EM corrections $\Psi^G$ in Eq.~\eqref{Psi} to the standard 3.5PN TaylorF2 phase for neutral BBHs $\Psi^{(F2)}_{3.5}(f)$ in  Eq.~(3.18) in \cite{PhysRevD.80.084043} and subtract the duplicated neutral 1PN contribution:
\begin{align}
\hat{\Psi}^G_+(f)=\Psi^G_+(f)+\Psi^{(F2)}_{+3.5}(f)-\Psi^G_+(f,\kappa_1=\kappa_2=0),
\end{align}
{where $\Psi_+^G(f,\kappa_1=\kappa_2=0)$ is the neutral part of $\Psi^G$ at 1PN order in Eq.~\eqref{Psi}, serving to subtract duplicate 1PN order contributions.} 

Compared with the standard 3.5PN waveform for neutral BBHs, this waveform model introduces two additional charge-related parameters, $\kappa_1$ and $\kappa_2$. If only leading electric-dipole radiation is retained, only the combination $\xi_1^2$ can be constrained.  Including electric multipole radiation through 1PN order introduces the additional combinations $\zeta_i$ and $\xi_i$ in Eq.~\eqref{Psi} as shown in the right panel of Fig.~\ref{GW_Q}, which can partially reduce the degeneracy between $\kappa_1$ and $\kappa_2$.

For Bayesian inference, we adopt the prior on the chirp mass chosen to be uniform in the range of $\mathcal{M}_c\in[5,15]~M_\odot$. The prior on the mass ratio chosen to be uniform in the range of $q\in[0.125,1]$. The priors on the charge parameters chosen to be uniform in the range of $\kappa_1\in[-0.8,0.8]$ and $\kappa_2\in[0,0.8]$. Because the waveform is invariant under $(\kappa_1,\kappa_2)\to(-\kappa_1,-\kappa_2)$, we restrict $\kappa_2\geq0$ to remove the duplicated posterior mode. This convention preserves both same-sign $(\mathcal{Q} > 0)$ and opposite-sign $(\mathcal{Q} < 0)$ configurations. The luminosity-distance prior is uniform in comoving volume over $ D \in[50,~2000]$ Mpc. We set the GW frequency $f\in[20~{\rm Hz},~f_{\rm ISCO}]$, where the waveform is terminated at the parameter-dependent ISCO frequency. For the component spins, we adopt the AlignedSpin prior implemented in B{\footnotesize ILBY}. The spin magnitudes are assigned independent uniform priors.
For other parameters, we use the same priors in the parameterized post-Einsteinian waveform model presented in Abbott et al \cite{PhysRevX.9.031040}.

The constraints on $\kappa_1$ and $\kappa_2$ obtained from GW$240925$, GW$241102$, and GW$250119$ are summarized in Table~\ref{constrain}. The table reports the signed component charge-to-mass ratios $\kappa_1$ and $\kappa_2$, their product $\mathcal{Q}$, and the component masses $m_1$ and $m_2$, and the chirp mass $\mathcal{M}_c$. The 68\% credible intervals for $\kappa_1$ span zero for GW240925 $(\kappa_1\in[-0.140,0.018])$, GW241102 $(\kappa_1\in[-0.143,0.008])$, and GW250119 $(\kappa_1\in[-0.070,0.216])$.  Because $\kappa_2\geq 0$ by convention, these intervals include both $\mathcal{Q}<0$ and $\mathcal{Q}>0$ configurations, as well as $\mathcal{Q}=0$. The posterior medians of $\mathcal{Q}$ are negative for GW240925 and GW241102 and positive for GW250119. In our analysis, the charge parameters are correlated with the mass parameters, and allowing nonzero charges shifts the component-mass posteriors toward lower values relative to the neutral analysis. At the stated credible level, the corresponding bounds on the magnitude of the charge product are $|\mathcal{Q}|\lesssim 0.010, ~ 0.009,~ \text{and} ~0.042$ for GW240925, GW241102, and GW250119, respectively.  At the lower cutoff frequency $f_0=20$ Hz of LIGO–Virgo–KAGRA, for a binary total mass of $m=20 M_\odot$, we obtain $(\xi_{1}^E)^2=48/5\mathcal{Z}^{-4/3}( G m \pi f_0/c^3)^{2/3}\thicksim 0.33$. The approximation $\Psi^G$ is valid only for posterior samples that satisfy $\xi_1^2=(\kappa_1-\kappa_2)^2\ll(\xi_{1}^E)^2$ throughout the analyzed frequency band.

\section{Conclusion}

We have constructed GW waveforms for charged compact binaries in Einstein--Maxwell theory. We obtained the frequency-domain waveform with the stationary-phase approximation, incorporating electromagnetic corrections to the amplitude and to the Fourier phase. The electromagnetic sector produces a characteristic PN hierarchy. 

We derived both gravitational-quadrupole domainated and electric-dipole dominated expansions of the phase and coalescence time, corresponding respectively to the gravitational quadrupole dominated and electromagnetic dipole dominated regimes. Our numerical results show that charge affects the accumulated phase, the inspiral time, and the ISCO cutoff frequency. Opposite-sign charges lead to a stronger phase shift, while same-sign charges reduce the effective attraction. For equal charge-to-mass ratios, the leading electric-dipole radiation vanishes, making higher-order electromagnetic radiation essential.  { As shown in Fig. \ref{EMPN}, there is a transition region where both the gravitational-quadrupole dominated and the electric-dipole dominated phases are invalid. It is therefore necessary to derive an analytic phase expression uniformly valid across these two regimes. This work is currently in preparation.}

By incorporating EM corrections through 1PN order, we reduce the degeneracy between the two component charge-to-mass ratios and retain higher-order EM contributions that remain nonzero for like-signed charges. We also applied the charged waveform to Bayesian inference for selected binary-black-hole events and introduced two additional charge-to-mass-ratio parameters, $\kappa_1$ and $\kappa_2$. The Bayesian analysis yields posterior medians and 68\% credible intervals for the component charge-to-mass ratios and $\mathcal{Q}$. The corresponding component masses are also reported. The resulting constraints illustrate the potential of inspiral GWs to probe electric charge, although more accurate waveform models are still needed.

This work represents a step toward analytic charged-compact-binary waveform models beyond the leading electric-dipole approximation. Future work should include higher-PN radiation effects, eccentricity, magnetic or hidden-sector charges. Matching the inspiral model to numerical-relativity simulations and charged merger-ringdown waveforms will be crucial for building complete inspiral-merger-ringdown models for future GW tests of black-hole charge.

\section{ACKNOWLEDGMENTS}
T.L. is supported by the Talent Introduction Reform and Development Subsidy of Yangtze University under Grant No. \,80210038. This work is supported in part by the National Natural Science Foundation of China under Grant No. 12475067 and No. 12235019, in part by the Natural Science Foundation of Hunan Province Grant No. 2026JJ60110.

\begin{widetext}
\appendix
\section{Parameters of orbital angular frequency}\label{App.A}
This Appendix lists the complete EM coefficients $\mathcal{C}_i$ and $\mathcal{D}_i$ entering the orbital-frequency evolution $\dot{\omega}$ in Eq. \eqref{Domega}; these coefficients were derived in our previous work \cite{zhang2026postnewtoniandynamicschargedcompact}
\begin{align}
    \mathcal{C}_0=&\xi _1^2 \left(\frac{5 \zeta _0}{3}+\frac{5 \eta }{3}-1\right)+\frac{24 \zeta _2^2}{5}-2 \xi _0 \xi _1-8 \xi _1 \xi _2,\\
     \mathcal{C}_1=&\xi _1^2 \left(-\zeta _0+4 \eta +6\right)-\frac{1}{5} 48 \zeta _2^2+2 \xi _0 \xi _1+\xi _1 \left(8 \xi _2-2 \xi _3\right)-\frac{192}{5},\\
     \mathcal{C}_2=& \xi _1^2\left(\frac{4 \zeta _0}{3}-11 \eta -\frac{28}{3}\right)+\frac{24 \zeta _2^2}{5}+2 \xi _1 \xi _3+\frac{96}{5},\\
     \mathcal{C}_3=&\xi _1^2\left(\frac{16 \eta }{3}+\frac{1}{3}\right) , \\
    \mathcal{D}_0=&\xi _1^2 \left(\zeta _0 \left(\frac{11 \eta }{6}-\frac{13}{6}\right)-\frac{16 \zeta _0^2}{9}+\frac{\eta ^2}{2}+\eta +\frac{1}{2}\right)+\xi _0 \xi _1 \left(\frac{4 \zeta _0}{3}-\frac{5 \eta }{3}+1\right)+\xi _1 \xi _2 \left(\frac{16 \zeta _0}{3}-\frac{20 \eta }{3}+4\right) \notag\\
    &+\zeta _2^2\left(\frac{36 \eta }{5}-12\right)+\left(-\frac{96 \zeta _3}{5}-\frac{16 \zeta _4}{7}\right) \zeta _2+\zeta _0 \left(6 \zeta _2^2+\frac{16}{5}\right)+\left(\frac{7314 \eta }{175}-\frac{3657}{350}\right) \xi _3^2,\\
    \mathcal{D}_1=&\xi _1^2 \left(\zeta _0 \left(\frac{103}{6}-\frac{7 \eta }{6}\right)+\frac{10 \zeta _0^2}{9}+\frac{11 \eta ^2}{3}+3 \eta +10\right)+\xi _0 \xi _1 \left(-\frac{4 \zeta _0}{3}-\frac{\eta }{3}-13\right)+\xi _1\xi _2 \left(-\frac{16 \zeta _0}{3}-\frac{4 \eta }{3}-52\right)\notag\\
    &+\xi _1\xi _3 \left(\frac{4 \zeta _0}{3}-\frac{5 \eta }{3}+1\right)+\zeta _2^2 \left(\frac{228}{5}-\frac{12 \eta }{5}\right)+\left(\frac{288 \zeta _3}{5}+\frac{64 \zeta _4}{7}\right) \zeta _2+\zeta _0 \left(\frac{176}{5}-14 \zeta _2^2\right)\notag\\
    &+\left(\frac{7314}{175}-\frac{29256 \eta }{175}\right) \xi _3^2+\frac{1584 \eta }{5}+\frac{6952}{35},\\
    \mathcal{D}_2=&\xi _1^2 \left(\zeta _0 \left(-\frac{17 \eta }{2}-\frac{143}{6}\right)+\frac{4 \zeta _0^2}{3}-\frac{14 \eta ^2}{3}+\frac{15 \eta }{2}-22\right)+\xi _0 \xi _1 \left(-\frac{4 \zeta _0}{3}+11 \eta +\frac{73}{3}\right)+\xi _1\xi _2 \left(-\frac{16 \zeta _0}{3}+44 \eta +\frac{292}{3}\right)\notag\\
    &+\xi _1\xi _3 \left(-\frac{4 \zeta _0}{3}-\frac{\eta }{3}-13\right)+\zeta _2^2 \left(-\frac{244 \eta }{5}-\frac{402}{5}\right)+\left(-\frac{288 \zeta _3}{5}-\frac{96 \zeta _4}{7}\right) \zeta _2+\left(\frac{66 \zeta _2^2}{5}+\frac{96}{5}\right) \zeta _0\notag\\
    &+\left(\frac{43884 \eta }{175}-\frac{10971}{175}\right) \xi _3^2-\frac{3264 \eta }{5}-\frac{2220}{7},\\
    \mathcal{D}_3=&\xi _1^2 \left(\zeta _0 \left(\frac{37 \eta }{2}+\frac{39}{2}\right)-\frac{14 \zeta _0^2}{9}-21 \eta ^2-\frac{343 \eta }{6}-\frac{13}{2}\right)+\xi _0 \xi _1 \left(\frac{4 \zeta _0}{3}-\frac{43 \eta }{3}-\frac{41}{3}\right)+\xi _1 \xi _2 \left(\frac{16 \zeta _0}{3}-\frac{172 \eta }{3}-\frac{164}{3}\right)\notag\\
    &+\xi _1\xi _3 \left(-\frac{4 \zeta _0}{3}+11 \eta +\frac{73}{3}\right)+\zeta _2^2 \left(\frac{444 \eta }{5}+\frac{366}{5}\right)+\left(\frac{96 \zeta _3}{5}+\frac{64 \zeta _4}{7}\right) \zeta _2+\left(-\frac{1}{5} 42 \zeta _2^2-\frac{208}{5}\right) \zeta _0\notag\\
    &+\left(\frac{7314}{175}-\frac{29256 \eta }{175}\right) \xi _3^2+\frac{2928 \eta }{5}+\frac{7456}{35},\\
    \mathcal{D}_4=& \xi _1^2 \left(\zeta _0 \left(-16 \eta -\frac{34}{3}\right)+\frac{8 \zeta _0^2}{9}+\frac{281 \eta ^2}{6}+\frac{457 \eta }{6}+19\right)+\xi _0 \xi _1\left(\frac{16 \eta }{3}+\frac{4}{3}\right) +\xi _1\xi _2\left(\frac{64 \eta }{3}+\frac{16}{3}\right)\notag\\
    &+\xi _1 \xi _3 \left(\frac{4 \zeta _0}{3}-\frac{43 \eta }{3}-\frac{41}{3}\right)+\zeta _2^2 \left(-\frac{288 \eta }{5}-\frac{138}{5}\right)-\frac{16 \zeta _4 \zeta _2}{7}+\left(\frac{16 \zeta _2^2}{5}+\frac{64}{5}\right) \zeta _0\notag\\
    &+\left(\frac{7314 \eta }{175}-\frac{3657}{350}\right) \xi _3^2-216 \eta -\frac{398}{7},\\
     \mathcal{D}_5=& \xi _1^2 \left(\zeta _0 \left(\frac{16 \eta }{3}+\frac{2}{3}\right)-\frac{100 \eta ^2}{3}-\frac{191 \eta }{6}-\frac{5}{6}\right)+\zeta _2^2 \left(\frac{64 \eta }{5}+\frac{6}{5}\right)+\left(\frac{16 \eta }{3}+\frac{4}{3}\right) \xi _1 \xi _3+\frac{96 \eta }{5}+\frac{24}{5},\\
    \mathcal{D}_6=&\xi _1^2\left(8 \eta ^2+\frac{4 \eta }{3}-\frac{1}{6}\right).
\end{align}
Assuming small charges and retaining terms through $\mathcal{O}(\kappa^2)$, Eq.~\eqref{Domega} reduces to
\begin{align}\label{Domega_C}
    \dot{\omega}=&\frac{c^6\eta \mathrm{x}^{9/2}}{G^2 m^2}\bigg\{2\xi_1^2+\frac{\mathrm{x}}{\mathcal{Z}^{4/3}}\bigg[\frac{96}{5}+\xi _1^2 \left(\frac{5 \eta }{3}-1\right)+\frac{24 }{5}\zeta _2^2-2 \xi _0 \xi _1-8 \xi _1 \xi _2-\frac{192}{5}\mathcal{Q}\bigg]\notag\\
    &+\frac{\mathrm{x}^2}{\mathcal{Z}^{8/3}}\bigg[-\frac{264 }{5}\eta-\frac{1486}{35}+\xi _1^2 \left(\frac{\eta ^2}{2}+\eta +\frac{1}{2}\right)+\xi _0 \xi _1 \left(-\frac{5 \eta }{3}+1\right)+\xi _1 \xi _2 \left(-\frac{20 \eta }{3}+4\right)+\zeta _2^2\left(\frac{36 \eta }{5}-12\right) \notag\\
    &+\left(-\frac{96 \zeta _3}{5}-\frac{16 \zeta _4}{7}\right) \zeta _2+\frac{16}{5}\zeta _0 +\left(\frac{7314 \eta }{175}-\frac{3657}{350}\right) \xi _3^2+\left(\frac{1584 \eta }{5}+\frac{6952}{35}\right)\mathcal{Q}\bigg]\bigg\}+\mathcal{O}(c^{-9}).
\end{align}
The coefficients above contain the complete EM multipole corrections to  $\dot{\omega}$ through the 1PN order.  In the neutral limit $\kappa_1=\kappa_2=0$, all charge combinations vanish and $\mathcal{Z}=1$, so Eq.~\eqref{Domega_C} reduces to the standard neutral-binary result. 
\end{widetext}

\bibliography{BIB.bib}
\bibliographystyle{apsrev4-1}
\setcitestyle{numbers,maxcitenames=3}

\end{document}